\documentclass[a4paper,twocolumn,11pt]{quantumarticle}
\pdfoutput=1

\usepackage{graphicx}
\usepackage{bm}
\usepackage{amsmath,amssymb,amsfonts,xfrac}
\usepackage{mathrsfs}
\usepackage[numbers,sort&compress]{natbib}
\usepackage{footnote}
\usepackage{braket}
\usepackage{xcolor}
\usepackage{hyperref}
\usepackage{orcidlink}

\begin{document}

\title{Charge self-consistent DFT+QDMFT framework with a Krylov-based quantum impurity solver}

\author[1,2]{M. Karabin\orcidlink{0000-0003-0081-8497}}
\email{mariia.karabin@mtsu.edu}
\author[3]{T. Sohail \orcidlink{0000-0001-7567-6417}}
\author[3]{E. A. Coello P\'erez\orcidlink{0000-0001-6074-0711}}
\email{coellopereea@ornl.gov}
\author[3]{S. Ghosh\orcidlink{0000-0003-3800-5264}}
\author[3]{M. Sajjan\orcidlink{0000-0001-7436-5422}}
\author[1,2]{H. Terletska\orcidlink{0000-0001-6215-9664}}
\author[3]{M. Eisenbach\orcidlink{0000-0001-8805-8327}}
\affil[1]{Department of Physics and Astronomy, Middle Tennessee State University, Murfreesboro, TN, USA}
\affil[2]{Quantum Research Interdisciplinary Science and Education (QRISE) Center, Middle Tennessee State University, Murfreesboro, TN, USA}
\affil[3]{National Center for Computational Sciences, Oak Ridge National Laboratory, Oak Ridge, TN, USA}


\begin{abstract}
We integrate a quantum impurity solver into a charge self-consistent density functional theory (DFT) plus quantum dynamical mean-field theory framework. The solver optimizes adaptive circuits to approximate the ground state of the impurity model and its particle and hole excitations, and reconstructs the impurity Green's function from their Hamiltonian moments. Higher-order moments are predicted from low-order ones using a machine-learning model, while regularized optimization mitigates violations of causality in the resulting self-energy. The latter is used to update the charge density until self-consistency is achieved. We demonstrate the framework through body-centered cubic lithium calculations and compare the results with a reference one using a continuous-time quantum Monte Carlo solver. These calculations converge with root-mean-square charge-density residuals below $\boldsymbol{2\times10^{-3}}$ and yield a nearly vanishing impurity self-energy, consistent with the noninteracting limit used for weakly correlated lithium.
\end{abstract}

\footnotetext{Notice: This manuscript has been authored by UT-Battelle, LLC, under contract DE-AC05-00OR22725 with the US Department of Energy (DOE). The US government retains and the publisher, by accepting the article for publication, acknowledges that the US government retains a nonexclusive, paid-up, irrevocable, worldwide license to publish or reproduce the published form of this manuscript, or allow others to do so, for US government purposes. DOE will provide public access to these results of federally sponsored research in accordance with the DOE Public Access Plan (\href{https://www.energy.gov/doe-public-access-plan}{https://www.energy.gov/doe-public-access-plan}).}

\section{Introduction}

Predictive simulations of strongly correlated materials require an accurate treatment of both material-specific electronic structure and local many-body interactions. Density functional theory (DFT) provides an efficient first-principles description of the electronic structure, but conventional exchange-correlation approximations can fail when electron localization and dynamical correlations become important. Dynamical mean-field theory (DMFT) addresses this limitation by mapping the correlated lattice problem onto a self-consistent impurity problem, providing access to frequency-dependent local correlations and spectral properties~\cite{Georges1996DMFT}. Combined DFT+DMFT methods therefore bridge material-specific electronic structure and local many-body physics, providing a first-principles framework for systems that cannot be described adequately by conventional DFT or by isolated model Hamiltonians alone~\cite{Kotliar2006DFTDMFT}.

DMFT accuracy and computational cost are governed by the solution of the auxiliary Anderson impurity model (AIM). Established impurity solvers provide complementary capabilities. Continuous-time quantum Monte Carlo (CTQMC) offers a treatment of broad classes of impurity problems at finite temperature~\cite{Werner2006CTQMC, Gull2011CTQMC}, whereas Hamiltonian-based approaches, such as exact diagonalization (ED), provide direct access to many-body eigenstates and real-frequency spectral information at the cost of an exponential growth of the many-body Hilbert space with system size~\cite{Caffarel1994ED}. This unfavorable scaling has motivated the development of quantum impurity solvers, in which the interacting fermionic degrees of freedom are encoded directly into qubit operators. Quantum dynamical mean-field theory (QDMFT) has consequently emerged as a promising framework incorporating quantum impurity solvers into DMFT calculations and addressing correlated-electron problems on quantum processors~\cite{Bauer2016Hybrid, Rungger2020QuantumDMFT, Ehrlich2025DMFT, Keen2020QuantumDMFT}.

While early QDMFT studies focused on the solution of model impurity problems~\cite{Jamet2022}, recent efforts have extended the approach to material-specific DFT+QDMFT calculations. In particular, a framework using a quantum impurity solver based on the variational quantum eigensolver (VQE) and quantum equation of motion algorithms~\cite{Peruzzo2014VQE, Rizzo2022QEOM} has recently been demonstrated~\cite{Selisko2025RealMaterials}. Integrating a quantum solver within self-consistent DFT+QDMFT calculations introduces additional challenges beyond those encountered in idealized benchmarks of the solvers themselves. The frequency-dependent hybridization function generated by the lattice calculation must be discretized into a finite set of bath degrees of freedom, sufficiently large to provide an accurate representation of the lattice problem while remaining compatible with the resources and limitations of available quantum hardware. At the same time, the resulting quantum calculation must remain consistent with the energy and Green's function conventions of the surrounding DFT+DMFT framework.

In previous work~\cite{Karabin2026QDMFT}, we developed a quantum AIM solver (QAIMS) for particle-hole-symmetric models. The solver optimized quantum circuits to prepare the ground state of the model and its particle and hole excitations, from which the ground-state energy and Hamiltonian moments of the excited states were estimated. These quantities were then used to approximate the impurity Green's function through its Lanczos continued fraction. The present work goes beyond the standalone impurity problem by embedding QAIMS into a self-consistent, material-specific DFT+QDMFT calculation. At each iteration, the lattice calculation defines a local impurity problem through the impurity level and frequency-dependent hybridization function. The latter is discretized to construct an explicit AIM, which is solved by QAIMS to obtain the impurity Green's function. The corresponding frequency-dependent impurity self-energy is then returned to the lattice calculation to generate the impurity problem for the next iteration, closing the self-consistency loop.

To assess the performance of QAIMS within a self-consistent DFT+QDMFT framework, we considered body-centered cubic (BCC) lithium. This material choice is motivated by two considerations. First, setting the Coulomb interaction to zero, a reasonable approximation for a weakly correlated material such as lithium, provides a simple baseline for testing the DFT+QDMFT implementation. 
In this noninteracting limit, any impurity solver should yield a vanishing self-energy, providing a benchmark for assessing errors arising from approximations and numerical procedures inherent to the solver. Furthermore, since the self-energy vanishes, DMFT should leave the charge density unchanged relative to the corresponding DFT solution. Therefore, any residual changes in the charge density provide a measure of how numerical and solver errors propagate through the self-consistency cycle. Second, from a materials physics perspective, electronic correlations in lithium may become increasingly relevant under compression or structural modifications, as changes in the band structure can enhance the relative importance of electron–electron correlations~\cite{matsuoka_t_2009, iyakutti_k_2004, christensen_ne_2001}. DFT+DMFT provides a framework for describing these effects, thereby enabling the investigation of correlation-driven changes in lithium's electronic structure beyond standard DFT.

The remainder of this paper is organized as follows. Section~\ref{sec:methods} describes the self-consistent DFT+DMFT framework and the methods used to discretize the lattice hybridization function and calculate the impurity Green's function. Section~\ref{sec:results} presents DFT+QDMFT results for BCC lithium using three- and five-bath-site discretizations. We assess the accuracy of the discrete-bath representation, the resulting impurity self-energy, and the impact of sampling noise on the convergence of DFT+QDMFT calculations. Finally, Section~\ref{sec:conclusions} summarizes the main findings and discusses their implications.

\section{Methods}
\label{sec:methods}
In this section, we describe the DFT+QDMFT framework, the methods used by QAIMS to solve the impurity problem, and the integration of the solver into the workflow. We first outline the self-consistency procedure and the construction of the AIM from the lattice hybridization function. Then, we describe the adaptive routines preparing the relevant quantum states and reconstructing the impurity Green's function from Hamiltonian moments, including the use of machine-learning (ML) to predict higher-order ones.

\subsection{The self-consistent loop}

We consider a DFT+QDMFT workflow in which the material-specific lattice problem is treated using Questaal~\cite{Pashove2020Questaal}, a suite of first-principles electronic-structure methods. Starting from a DFT calculation within the local density approximation that uses the von Barth-Hedin exchange-correlation functional~\cite{vonBarth1972}, a correlated subspace is constructed by projecting the Bloch eigenstates, denoted by $\nu$ and $\nu'$, onto selected localized orbitals, denoted by $L$ and $L'$. Within the correlated subspace, DMFT approximates the many-body self-energy as local and maps the lattice problem onto an effective AIM, which is solved self-consistently with the lattice electronic structure~\cite{Georges1996DMFT, Kotliar2006DFTDMFT}. The calculation comprises an inner impurity-lattice self-consistency loop and an outer loop that updates the electronic density and the associated DFT potential~\cite{Haule2010DFT_DMFT}. We first describe the inner loop at outer iteration $n$, for which the fixed charge density is $\rho^{[n]}$.

At the $m$-th inner iteration, the impurity self-energy $\Sigma^{[m-1]}_{\rm imp}(z)$, solution of the AIM at the preceding iteration, is embedded into the lattice problem. To do so, it is transformed from the correlated-orbital basis to the Bloch basis via projectors $P_{L\nu}(\mathbf{k})\equiv\braket{L|\nu\mathbf{k}}$, which connect the Bloch states $\ket{\nu\mathbf{k}}$ to the correlated orbitals $L$
\begin{equation}
    \Sigma^{[m-1]}_{\rm emb,\nu\nu'}(\mathbf{k},z) = \sum_{LL'} P_{\nu L}^{\dagger} \Sigma^{[m-1]}_{{\rm imp},LL'}(z) P_{L'\nu'} .
\end{equation}
For brevity, we have suppressed the explicit arguments of previously defined objects in the equation above. In what follows, the arguments of previously defined quantities are similarly omitted. The lattice Green's function incorporates the embedded self-energy into the noninteracting DFT Green's function,
\begin{equation}
    G^{[m]}(\mathbf{k},z) = \left\{ \left[G^{\mathrm{DFT}}(\mathbf{k},z) \right]^{-1} - \Sigma^{[m-1]}_{\rm emb} \right\}^{-1}.
\end{equation}
The impurity self-energy is initialized to $\Sigma^{[0]}_{\rm imp}=0$, so that the initial lattice Green's function coincides with the noninteracting DFT Green's function.

Projecting the lattice Green's function onto the localized orbitals and integration over the Brillouin zone yield the local Green's function
\begin{equation}
    G^{[m]}_{\mathrm{loc},LL'}(z) = \sum_{\mathbf{k},\nu\nu'} w_{\mathbf{k}} P_{L\nu} G^{[m]}_{\nu\nu'} P_{\nu'L'}^{\dagger} ,
\label{eq:gloc}
\end{equation}
where $w_{\mathbf{k}}$ denotes the Brillouin-zone weight.

The local Green's function and the local self-energy determine the  Weiss field of the effective impurity problem
\begin{equation}
    \mathcal{G}_{0}^{[m]}(z) = \left[ \left(G^{[m]}_{\rm loc}(z) \right)^{-1} + \Sigma_{\rm imp}^{[m-1]}(z) \right]^{-1} ,
\label{eq:weiss}
\end{equation}
which can be expressed in terms of the hybridization function $\Delta^{[m]}(z)$ characterizing the coupling between the effective impurity and its surrounding bath. The Weiss field is the noninteracting propagator of the auxiliary impurity problem. It contains the dynamical coupling to the surrounding lattice before the local interaction is solved explicitly. For a single correlated orbital, the Weiss field takes the form 
\begin{equation}
    \mathcal{G}^{[m]}_{0}(z) = \left[z - \epsilon^{[m]}_{\rm imp} - \Delta^{[m]} \right]^{-1} ,
\label{eq:g0}
\end{equation}
where $\epsilon_{\rm imp}$ denotes the energy of the impurity site. The chemical potential and double-counting shifts entering this quantity are treated consistently with the lattice calculation.

The single-impurity AIM Hamiltonian is
\begin{equation}
    H_{\mathrm{AIM}} = H_{\mathrm{imp}} + H_{\mathrm{bath}} + H_{\mathrm{hyb}} ,
\label{eq:haim}
\end{equation}
with
\begin{gather}
    H_{\mathrm{imp}} = \sum_{\sigma} \epsilon_{\mathrm{imp}} d_{\sigma}^{\dagger}d_{\sigma} + U n_{d\uparrow}n_{d\downarrow} ,\nonumber\\
    H_{\mathrm{bath}} = \sum_{k\sigma} \epsilon_k c_{k\sigma}^{\dagger}c_{k\sigma} ,\nonumber\\
    H_{\mathrm{hyb}} = \sum_{k\sigma} V_k \left( d_{\sigma}^{\dagger}c_{k\sigma} + c_{k\sigma}^{\dagger}d_{\sigma} \right).
\end{gather}
Here $U$ is the local Coulomb interaction, $\epsilon_k$ denotes the energies of the bath sites, and $V_k$ are the impurity-bath coupling strengths.

The impurity Green's function $G^{[m]}_{\rm imp}$, obtained after solving the $m$-th iteration impurity problem, determines the $m$-th iteration impurity self-energy through Dyson's equation, 
\begin{equation}
    \Sigma_{\rm imp}^{[m]}(z) = z - \epsilon^{[m]}_{\rm imp} - \Delta^{[m]} - \left( G_{\rm imp}^{[m]} \right)^{-1} .
\label{eq:dyson_sigma}
\end{equation}
Feeding back this impurity self-energy into the lattice calculation closes the inner impurity-lattice loop. The inner convergence is monitored using the root-mean-square (RMS) change in the self-energy and the relative change in the lattice-derived hybridization
\begin{gather}
    \epsilon_\Sigma = \sqrt{\frac{1}{N_{\rm RMS}}\sum_{n=0}^{N_{\rm RMS}-1} \left|\Sigma^{[m]}(i\omega_n)-\Sigma^{[m-1]}(i\omega_n)\right|^2} ,\nonumber\\
    \delta\Delta = \frac{\left\| \Delta^{[m]}(i\omega_n) - \Delta^{[m-1]}(i\omega_n) \right\|_2}{\max\left( \left\| \Delta^{[m-1]}(i\omega_n) \right\|_2, 10^{-14} \right)} ,
\end{gather}
where $N_{\rm RMS}$ determines the low-frequency window used in the RMS evaluation.

After convergence of the inner DMFT loop, the electronic density is reconstructed from the lattice Green's function, incorporating the frequency-dependent impurity self-energy. With the self-energy held fixed, the density is then iteratively relaxed until the RMS charge residual \(\epsilon_{\rho}\) falls below \(2\times10^{-3}\). The resulting converged density, \(\rho^{[n+1]}\), is used to reconstruct the DFT effective potential and update the lattice Hamiltonian. If the charge residual meets the convergence criterion after this density update, the outer DFT loop is considered converged. Otherwise, a new DMFT loop is initiated using the updated lattice Hamiltonian.  The outer loop uses the electronic density obtained from the interacting lattice Green's function to update the Kohn-Sham potential. The calculations are spin unpolarized, neglect spin-orbit coupling, and use a $\Gamma$-centered $16\times16\times16$ Brillouin-zone mesh. The lattice DMFT calculations use an inverse temperature of $\beta=50~\mathrm{eV}^{-1}$ and 2000 Matsubara frequency points.

\subsection{Finite-bath discretization}
\label{sec:fitting}

Each DMFT iteration yields a continuous hybridization function $\Delta$ on the fermionic Matsubara axis. However, our quantum impurity solver operates on AIMs with a finite number of bath orbitals. Therefore, each iteration's lattice-derived hybridization function requires a discrete representation in terms of a finite number of bath energies $\epsilon_k$ and impurity-bath coupling strengths $V_k$. For a bath containing $N_b$ orbitals, this representation is written as
\begin{equation}
    \Delta_{\mathrm{dis}}(i\omega_n) = \sum_{k=1}^{N_b} \frac{\left|V_k\right|^2}{i\omega_n - \epsilon_k} .
\end{equation}
Together with the impurity energy $\epsilon_{\rm imp}$ and interaction strength $U$, the discrete bath parameters define the finite AIM subsequently solved by the quantum impurity solver.

Fitting the hybridization function is an important numerical step in DMFT, for which optimization methods based on semidefinite relaxation have been developed~\cite{MejutoZaera2020}. Here, the parameters defining the discrete bath are fitted to the hybridization function giving greater weight to lower Matsubara frequencies, as this regime is particularly important for the low-energy impurity physics entering the DMFT loop. Both the real and imaginary components are optimized simultaneously using the weighted nonlinear least-squares objective
\begin{equation}
    \chi^2 =
    \sum_{n=0}^{N_{\rm fit}-1}
    \frac{\left| \Delta_{\rm dis}(i\omega_n) - \Delta(i\omega_n) \right|^2}{|\omega_n|^{p}}
\end{equation}
where $N_{\rm fit}$ is the number of Matsubara points included in the fit and $p$ controls the emphasis placed on the low-frequency region. The bath fit uses the lowest $N_{\mathrm{fit}}=300$ Matsubara frequencies with weighting exponent $p=0.5$. The initial bath energies are distributed over an interval around the Fermi level, while the initial hybridization amplitudes are estimated from the low-frequency scale of the hybridization function.

The resulting finite AIM is then solved using the quantum impurity solver described in Section~\ref{sec:qaims}. Discretization of the hybridization function introduces a controlled modeling error that is distinct from errors associated with the quantum impurity solver itself. For consistency, the self-energy obtained from the quantum impurity solver is reconstructed from Dyson equation~\eqref{eq:dyson_sigma} using the discrete hybridization function that defines the finite AIM Hamiltonian.

\subsection{Adaptive ground-state preparation}
\label{sec:qaims}
\begin{figure}[!t]
    \centering
    \includegraphics[width=0.95\linewidth]{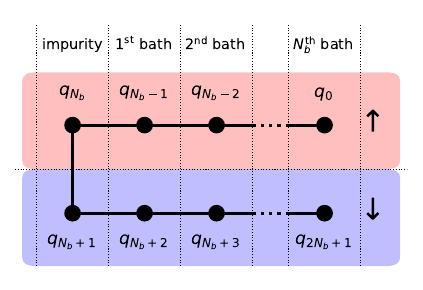}
    \caption{Orbital-to-qubit mapping used for the Jordan–Wigner transformation of an $N_b$-bath AIM}
    \label{fig:orbit-qubit-map}
\end{figure}

The impurity solver encodes the fermionic degrees of freedom of the AIM into a quantum register using the Jordan–Wigner transformation~\cite{Jordan1928Wigner}, following the orbital-to-qubit mapping in Fig~\ref{fig:orbit-qubit-map}.

\begin{figure}[!b]
    \centering
    \includegraphics[width=0.65\linewidth]{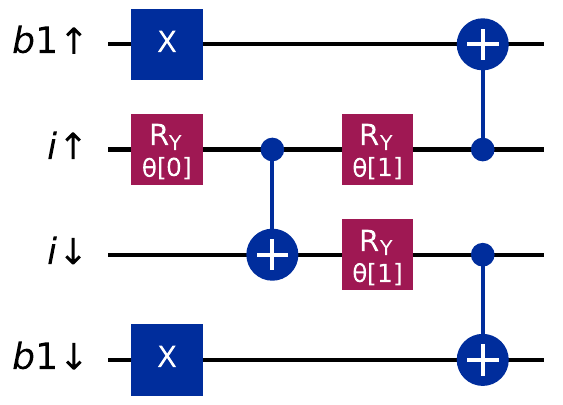}
    \caption{Quantum circuit preparing a spin-exchange-symmetric impurity state with occupation zero to two}
    \label{fig:imp_qc}
\end{figure}

In its current version, QAIMS employs adaptive derivative-assembled pseudo-Trotter VQE (ADAPT-VQE)~\cite{Grimsley2019ADAPT_VQE,Mukherjee2023} to construct and optimize quantum circuits preparing a state approximating the ground state of the AIM, as well as normalized particle and hole excitations of this optimized state. This construction starts from the circuit in Fig.~\ref{fig:imp_qc},
which prepares the impurity register in a spin-exchange-symmetric state
\begin{gather}
    \ket{\psi_{\rm imp}(\boldsymbol{\theta})} = A\ket{00} + B\ket{11} + C\left( \ket{01} + \ket{10} \right) ,\nonumber\\
    A\equiv \cos\frac{\theta_0}{2}\cos^2\frac{\theta_1}{2} + \sin\frac{\theta_0}{2}\sin^2\frac{\theta_1}{2} , \nonumber\\
    B\equiv \cos\frac{\theta_0}{2}\sin^2\frac{\theta_1}{2} + \sin\frac{\theta_0}{2}\cos^2\frac{\theta_1}{2} , \nonumber\\
    C\equiv \left( \cos\frac{\theta_0}{2} - \sin\frac{\theta_0}{2} \right) \frac{\sin\theta_1}{2} ,
\end{gather}
allowing QAIMS to describe ground states away from half filling. Entangling the impurity state with the populated first bath through CX gates yields a paramagnetic ground state with one particle in each spin sector. The spin-exchange symmetry of the impurity state is extended to the full AIM state by applying the same gates, with identical parameters, to the corresponding spin-up and spin-down registers throughout the remainder of the circuit construction.

\begin{figure}[!t]
    \centering
    \includegraphics[width=0.95\linewidth]{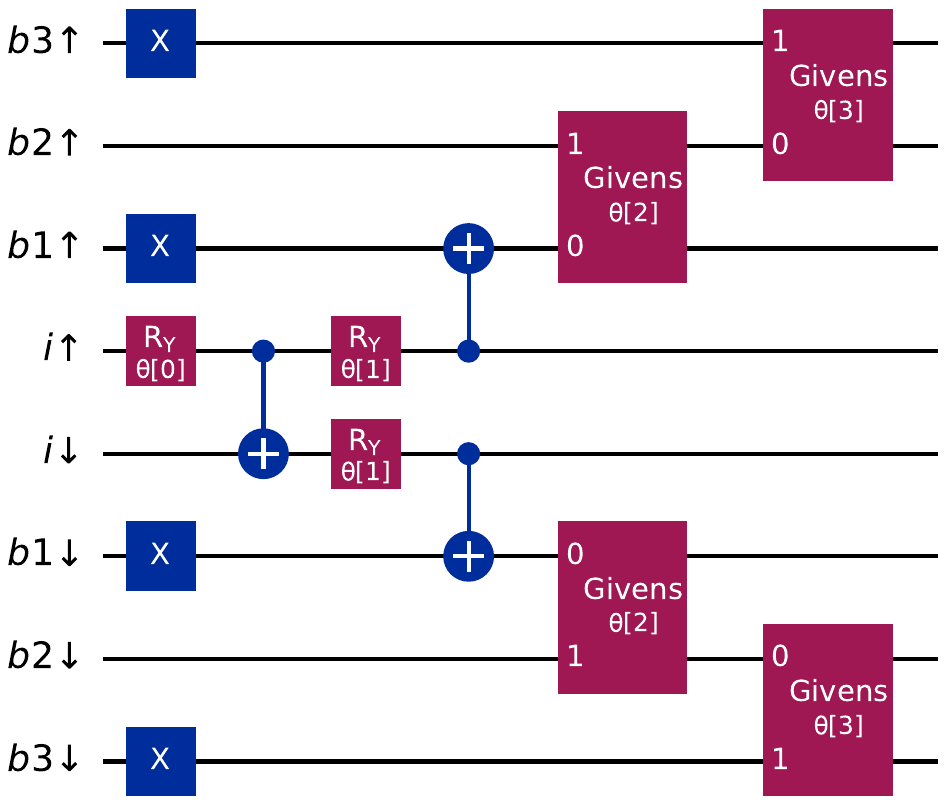}
    \caption{Initial circuit in the ADAPT-VQE routine solving a three-bath AIM. This circuit prepares a paramagnetic state with two electrons in each spin sector}
    \label{fig:initial_qc}
\end{figure}

To identify the particle-number sector containing the ground state, QAIMS evaluates the Hamiltonian expectation value for a set of candidate circuits, each enforcing a different bath-site occupation through a layer of X gates. The bath sites of the lowest-energy candidate are then entangled within each spin register using a ladder of Givens rotations, as illustrated in the example circuit in Fig.~\ref{fig:initial_qc}.
The resulting circuit serves as the starting point for the ADAPT-VQE algorithm.

QAIMS then employs the standard ADAPT-VQE procedure to iteratively expand the circuit using an operator pool consisting of Givens rotations between nearest-neighbor bath sites. At each iteration, the energy gradient associated with each operator in the pool is evaluated, and the operator with the largest gradient magnitude is appended to the circuit. The parameters of the resulting circuit are subsequently optimized to minimize the Hamiltonian expectation value. This procedure is repeated until the desired convergence criterion is reached. The optimized state $\ket{\psi_0}$ has energy $E_0$.

At each step of this adaptive construction, the circuit parameters are optimized using the COBYLA algorithm~\cite{Powell1994COBYLA} for up to 100 iterations. The initial measurement budget is set to 5000 shots per measured circuit for three-bath AIMs and 10000 shots for the five-bath ones.

\subsection{Impurity Green's function reconstruction}

\begin{figure*}[!t]
    \centering
    \includegraphics[height=0.245\linewidth]{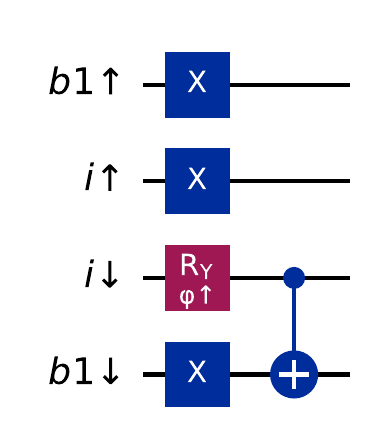}
    \includegraphics[height=0.245\linewidth]{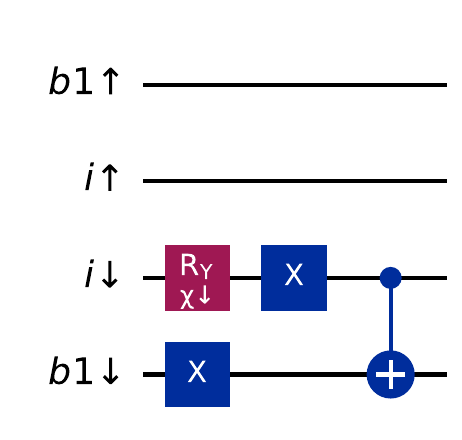}
    \includegraphics[height=0.245\linewidth]{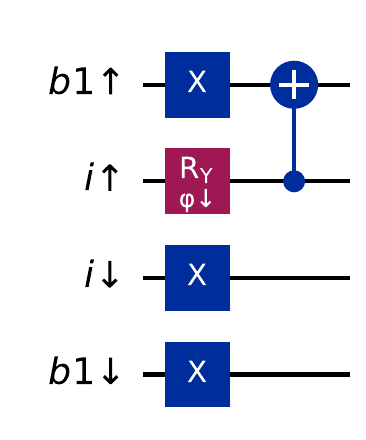}
    \includegraphics[height=0.245\linewidth]{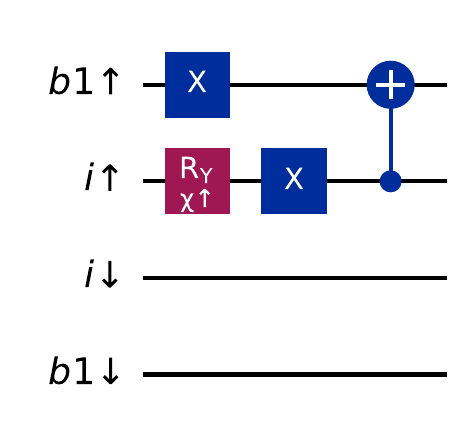}
    \caption{Quantum-circuit blocks used by QAIMS to prepare normalized particle and hole excitations. From left to right, the circuits prepare normalized spin-up particle, spin-down hole, spin-down particle, and spin-up hole excitations.}
    \label{fig:excited_qc}
\end{figure*}
The proposed ground-state circuit construction also allows us to readily prepare normalized particle and hole excitations of the optimized state, $\mathscr{N}_p d_\sigma^\dagger\ket{\psi_0}$ and $\mathscr{N}_h d_\sigma\ket{\psi_0}$, with normalization constants
\begin{gather}
    \mathscr{N}_p = \left(A^2 + C^2 \right)^{-1/2} ,\nonumber\\
    \mathscr{N}_h = \left(B^2 + C^2 \right)^{-1/2} .
\end{gather}
This is achieved by replacing the block in Fig.~\ref{fig:imp_qc} within the ground-state circuit with the corresponding blocks shown in Fig.~\ref{fig:excited_qc}. The angles in these blocks are given by
\begin{gather}
    \varphi_\uparrow = \varphi_\downarrow = 2\arctan\left(\frac{C}{A}\right) ,\nonumber\\
    \chi_\uparrow = \chi_\downarrow = 2\arctan\left(\frac{C}{B}\right) .
\end{gather}

Evaluating the Hamiltonian moments of the normalized excitations allows QAIMS to compute Lanczos coefficients from Hankel determinants~\cite{Witte1994LanczosCoeffs}. These coefficients define the impurity Green's function through its continued-fraction representation~\cite{GreeneDiniz2024, Haydock1972Recursion},
\begin{align}
    G_{\rm imp, \sigma}(z) &= \frac{1/\mathscr{N}_p^2}{z - \alpha_0^{p\sigma} + E_0 - \frac{(\beta_1^{p\sigma})^2}{z - \alpha_1^{p\sigma} + E_0 - \cdots}} \nonumber\\
    & + \frac{1/\mathscr{N}_h^2}{z + \alpha_0^{h\sigma} - E_0 - \frac{(\beta_1^{h\sigma})^2}{z + \alpha_1^{h\sigma} - E_0 - \cdots}} .
\end{align}
Here, the particle and hole Lanczos coefficients for spin $\sigma$ are calculated from the corresponding excitation sectors. Related moment-based approaches reconstruct Green's functions from particle and hole spectral moments, both within coupled-cluster theory~\cite{Backhouse2022} and using compact variational quantum circuits for impurity models~\cite{Sakurai2024}.

The impurity self-energy is then obtained from the Weiss field and impurity Green's function according to
\begin{equation}
    \Sigma_{\rm imp}(z) = z - \epsilon_{\rm imp} - \Delta_{\rm dis} - \left( G_{\rm imp} \right)^{-1} ,
\end{equation}
where the Weiss field is defined using the same discretized hybridization function that defines the finite AIM. This choice ensures consistency between the impurity Green's function and the noninteracting impurity propagator and reduces errors associated with representing the continuous hybridization function with a finite set of bath orbitals.

\subsection{Measurement shots allocation}

QAIMS distributes the shot budget for an expectation value among the circuits associated with the operator's commuting groups using a coefficient-weighted allocation scheme. Each group is assigned a weight
\begin{equation}
    w_g = \frac{\sum_{n\in {\rm group}}|c_n|}{\sum_n |c_n|},
\end{equation}
where $c_n$ is the coefficient of the $n$-th Pauli string. The shot budget, $N_s N_g$, where $N_s$ is the number of shots per circuit and $N_g$ is the number of commuting groups, is then allocated according to these weights, while imposing a minimum allocation of $N_s/5$ shots per group. Specifically, each group is first assigned the minimum number of shots, and the remaining shot budget is distributed proportionally to \(w_g\). This allocation prioritizes groups containing Pauli terms with larger aggregate coefficients while ensuring that every group receives sufficient measurements to contribute to the expectation-value estimate.

The shot budget for estimating the highest Hamiltonian moment is chosen to yield a sampling error comparable to that of $E_0$. Accordingly, the number of shots per commuting group is increased to $\tilde{N}_s = N_s\tilde{N}_p/N_p$, where $N_p$ and $\tilde{N}_p$ denote the number of Pauli strings in the Hamiltonian and in the highest measured Hamiltonian power, respectively.


\subsection{Higher-moment prediction}
\label{sec:ml_moments}

The cost of estimating Hamiltonian moments on a quantum computer increases rapidly with moment order, eventually becoming computationally prohibitive. Based on previous work's ML prediction of the impurity Green's function from model parameters~\cite{Arsenault2014}, QAIMS uses a neural network to predict higher-order Lanczos coefficients from $K$ known moments and physical descriptors of the AIM. The neural network consists of a permutation-invariant bath encoder followed by a residual multilayer perceptron. The encoder uses a DeepSets architecture~\cite{zaheer2017deep, soelch2019deep, ostdiek2021deep, zhang2019deep}, reflecting the invariance of the impurity problem under relabeling of the bath orbitals.

For each particle and hole sector $s$, the Hamiltonian is centered and rescaled such that the first normalized moments satisfy $\nu_0^s=1$, $\nu_1^s=0$, and $\nu_2^s=1$, and the first Lanczos coefficients become $\alpha_0^s=0$, $(\beta_1^s)^2=1$, and $\alpha_1^2=\nu_3^s$. The network predicts the remaining coefficients, extending the continued-fraction representation to $N_k$ Krylov states, using the AIM Hamiltonian parameters, $K$ known normalized moments, and approximate spectral information obtained from noninteracting, atomic, and cumulant-based references, enforcing nonnegative squared off-diagonal coefficients to ensure a physically admissible Lanczos continued fraction.

The network is trained by minimizing the physics-informed objective function
\begin{equation}
    \mathcal{L} = w_\theta \mathcal{L}_\theta + w_A \mathcal{L}_A + w_\mu \mathcal{L}_\mu + w_\delta \mathcal{L}_\delta + w_\Sigma \mathcal{L}_\Sigma ,
\end{equation}
with $w_\theta$, $w_A$, $w_\mu$, $w_\delta$, and $w_\Sigma$ scalar hyperparameters controlling the relative contributions of the five loss terms. Here, $\mathcal{L}_\theta$ penalizes errors in the predicted Lanczos coefficients, $\mathcal{L}_A$ measures differences between the reconstructed and reference spectral densities, and $\mathcal{L}_\mu$ penalizes errors in the reconstructed normalized Hamiltonian moments. The term $\mathcal{L}_\delta$ is a Gaussian negative log-likelihood that accounts for coefficient-dependent prediction uncertainties. The final term $\mathcal{L}_\Sigma$ penalizes discrepancies in a sector-resolvent-based self-energy proxy, constraining the accuracy of the inverse resolvents. Signed-log transformations are applied to the coefficient, moment, and self-energy losses to accommodate their broad numerical ranges. Together, these contributions guide the network toward accurate Lanczos coefficients while accounting for the spectral and Green's-function properties they determine.

Separate models predict moments for three- and five-bath AIMs. These models reconstruct moments up to the orders extending the continued fraction to six and eight Krylov states, respectively. Their training includes examples with three, four and five known moments, allowing each model to predict the remaining coefficients from different amounts of input information. Training uses AdamW with a peak learning rate of $2\times10^{-4}$, weight decay $10^{-5}$, and batch size 1024, for a maximum of 300 epochs. A ten-epoch learning rate warm-up is followed by cosine decay, and the model is selected using validation performance. 

\subsection{Causality-preserving moment optimization}
Errors in the estimated and predicted Hamiltonian moments, together with truncation of the Lanczos continued fraction, can lead to causality violations in the impurity self-energy. Although causal optimization has previously been used to correct Green's functions in DMFT calculations~\cite{Han2021}, QAIMS penalizes causality violations in the self-energy by refining the Hamiltonian moments through a regularized least-squares optimization with the objective function
\begin{gather}
    \mathcal{L} = \mathcal{L}_{\rm caus} + \mathcal{L}_{\rm dev}, \nonumber\\
    \mathcal{L}_{\rm caus}(\boldsymbol{\mu}) =
    \lambda_{\mathrm{caus}} \sum_n
    \left[
    \max\left(0,{\rm Im}[\Sigma_{\boldsymbol{\mu}}(i\omega_n)] \right)
    \right]^2, \nonumber\\
    \mathcal{L}_{\rm dev}(\boldsymbol{\mu}) =
    \lambda_{\mathrm{dev}} \sum_i 
    \left( \frac{\mu_i-\mu_{\rm orig,i}}{\sigma_i} \right)^2 ,
\end{gather}
where $\boldsymbol{\mu}_{\rm orig}$ and $\boldsymbol{\mu}$ are the original and refined moments, and $\Sigma_{\boldsymbol{\mu}}$ is the self-energy computed from the latter. The first term penalizes positive values of the imaginary part of $\Sigma_{\boldsymbol{\mu}}$ which violate the causality condition. The second one penalizes deviations from the original moments, with each deviation normalized by a scale factor chosen to allow moment adjustments of approximately five percent while discouraging larger deviations. The resulting regularization improves the causality of the computed self-energy while preserving the original moment estimates to within their expected uncertainty.

\subsection{Reference CTQMC calculation}

To benchmark the BCC lithium DFT+QDMFT calculations, we perform a reference one employing CTQMC to solve the AIM. Both solvers share the same lattice construction and DMFT feedback procedure.

In the reference calculation, the lattice-derived hybridization is supplied directly to the CTQMC impurity solver~\cite{Haule2007CTQMC}. Consequently, no finite-bath parameterization is introduced. 
This construction provides a continuous-bath reference for the complete self-consistent calculations. The CTQMC calculation uses an inverse temperature of $\beta=50~\mathrm{eV}^{-1}$ and 
$1\times10^7$ Monte Carlo steps per core for each impurity solution. The first $6.25\times10^5$ are used for equilibration, and measurements are recorded every 20 steps. The lowest 300 Matsubara frequencies are sampled directly, while higher frequencies are treated using the solver's high-frequency tail construction. 

\section{Results}
\label{sec:results}

The accuracy with which the discrete bath captures the lattice hybridization and QAIMS solves the impurity problem directly impact the convergence and reliability of DFT+QDMFT calculations, as errors in the discrete-bath hybridization and the impurity self-energy propagate to the lattice Green's function through the self-consistency cycle. In this section, we evaluate the accuracy of the finite-bath discretization, the quality of the variational approximation to the AIM ground state, the performance of the ML models in predicting higher-order Hamiltonian moments, and the convergence of the resulting self-consistent DFT+QDMFT calculations for BCC lithium.

\subsection{Finite-bath discretization accuracy}

\begin{figure}[!t]
    \centering
    \includegraphics[width=\linewidth]{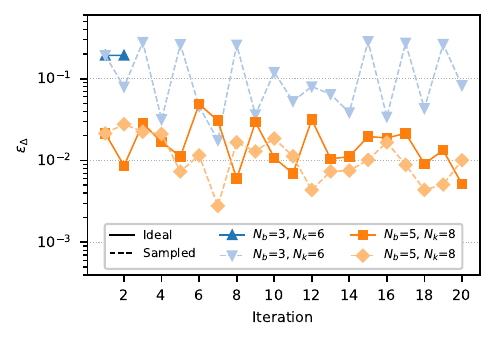}
    \caption{RMS error between the hybridization function and its discrete representation along a DMFT loop. Blue and orange markers represent errors for baths with three and five sites. Solid and dashed lines connect markers corresponding to QAIMS simulations without and with sampling noise considered.}
    \label{fig:delta_fit}
\end{figure}
The accuracy with which the discrete-bath hybridization captures the lattice one can be quantified using the RMS error
\begin{equation}
    \epsilon_{\Delta} = \sqrt{\frac{1}{N_{\rm RMS}}\sum_{n=0}^{N_{\rm RMS}-1} \left|\Delta_{\rm fit}(i\omega_n)-\Delta(i\omega_n)\right|^2} .
\end{equation}
We set $N_{\rm RMS}=N_{\rm fit}=300$, so the RMS fitting error is evaluated over the lowest 300 Matsubara frequencies. As shown in Fig.~\ref{fig:delta_fit}, the RMS error evaluated throughout the inner QDMFT loop, decreases by approximately one order of magnitude when the number of bath sites is increased from three to five. In the figure, markers connected by solid lines correspond to ideal QAIMS simulations, whereas those connected by dashed lines represent simulations that account for sampling.

\subsection{Ground-state fidelity}\label{sec:GS_fidelity}

Because the same variational parameters are used to prepare an approximation to the ground state and its normalized particle- and hole-excitations, inaccuracies in this approximation affect both the estimate for the ground-state energy and those for the Hamiltonian moments defining the Lanczos continued fraction.

\begin{figure}[!t]
    \centering
    \includegraphics[width=0.95\linewidth]{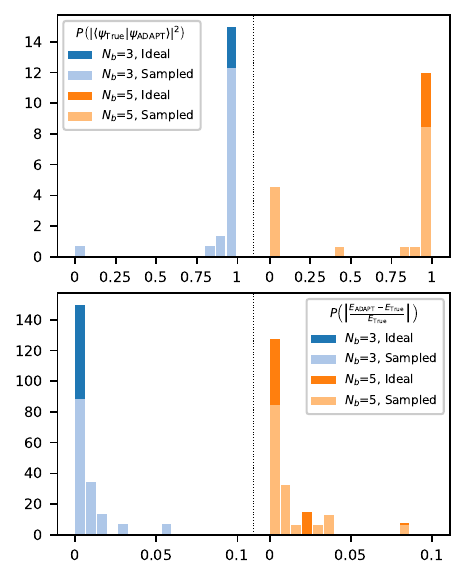}
    \caption{ADAPT-VQE performance within a DMFT loop. The top and bottom panels show the distributions for the overlap between the true ground state and its variational approximation, and the absolute relative difference between the true ground-state energy and its variational estimate. Blue and orange bars represent the distributions resulting from the solution of AIMs with three and five bath sites. Dark and light bars represent distributions corresponding to QAIMS simulations without and with sampling noise considered.}
    \label{fig:adapt_perf}
\end{figure}
\begin{figure*}[!t]
    \centering
    \includegraphics[width=0.95\linewidth]{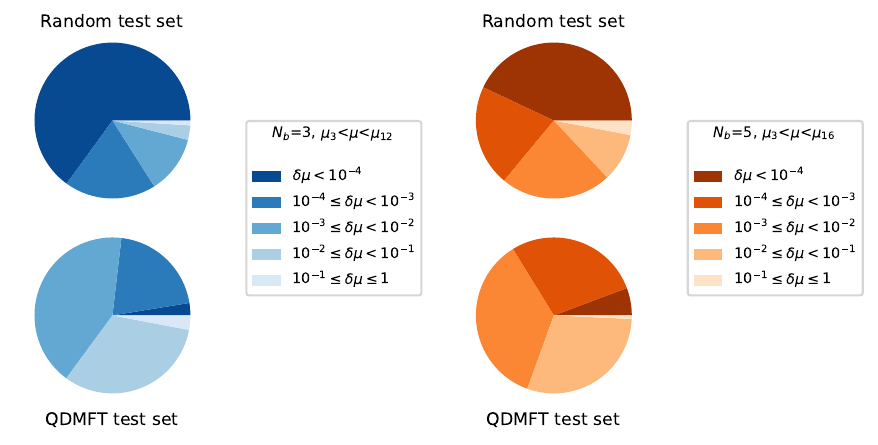}
    \caption{Absolute relative difference between predicted and true Hamiltonian moments. Blue and orange pies represent the relative moment differences for AIMs with three and five bath sites, respectively. The pies at the top were obtained from randomly generated AIM instances, while those at the bottom were obtained from the AIMs constructed within the QDMFT loop. Throughout the DFT+QDMFT calculations, more than 60 percent of the predicted moments yield differences below a percent}
    \label{fig:rel_mom_diff}
\end{figure*}
Throughout the DFT+QDMFT calculations, we use ED to obtain the true ground state of the AIM and assess the ability of QAIMS to find optimal parameters accurately approximating it. The distributions in the top panel of Fig.~\ref{fig:adapt_perf} show that QAIMS achieves ground-state overlaps above $0.9$ in most runs, even in the presence of sampling. The solver's performance deteriorates as the number of bath sites increases, reflecting the increased difficulty of approximating the ground state of larger AIMs. The distributions also show zero overlap cases, which indicate that the adaptive routine failed to identify the correct particle-number sector.

The corresponding distributions for the absolute relative difference between the true ground-state energy and its variational estimate are shown at the bottom of Fig.~\ref{fig:adapt_perf}. These differences are generally below five percent.

\subsection{Higher-moment model performance}

QAIMS estimates only the first three Hamiltonian moments through quantum measurement in each particle and hole sector and uses the ML models described in Section~\ref{sec:ml_moments} to predict higher moments needed to compute Lanczos coefficients up to the sixth and eighth Krylov states for the calculations solving AIMs with three and five bath sites, respectively. We assess the accuracy of these predictions by comparing them with true Hamiltonian moments obtained from the statevector representation of particle and hole excitations.

Figure~\ref{fig:rel_mom_diff} summarizes the performance of the trained models. The blue pie charts on the left show the distribution of absolute relative prediction errors, $\delta\mu$, with darker and lighter shades representing smaller and larger errors, respectively. The top pie chart summarizes predictions of excited-state moments for approximately 13,000 randomly generated AIM instances, using true values of the first three moments as input. The bottom pie chart shows predictions for the moments of the 20 AIMs constructed during the QDMFT loop, using estimates of the first three moments that account for sampling noise. While the model exhibits good predictive performance on the randomly generated test set, its accuracy decreases for the AIMs encountered throughout the QDMFT loop. This reduction is partly attributable to sampling noise, and is further compounded by the extreme bath energies and couplings that often arise from the three-bath discretization of the hybridization function, as these parameter values are less well represented in the randomly generated training instances, making the corresponding AIMs more challenging for the model to predict accurately.

The orange pie charts on the right show the corresponding prediction-error distributions for AIMs with five bath sites. For the randomly generated test set, the model exhibits lower predictive accuracy than the one for the three-bath case because it is trained to predict higher-order moments, which are generally more challenging to estimate. In contrast, its performance for AIMs encountered throughout the QDMFT loop improves relative to the three-bath case. This improvement can be attributed to the greater flexibility of the five-bath discretization in reproducing the hybridization function, which yields AIMs with more physically reasonable bath energies and coupling constants.

For both bath sizes, the predicted moments of the AIMs encountered throughout the QDMFT loop exhibit absolute relative errors below one percent in more than 60 percent of the cases.

\subsection{Charge-density convergence}

We evaluate the performance of QAIMS within the BCC lithium DFT+QDMFT calculations by examining the charge-density residual and the impurity self-energy. The calculations converge to the prescribed tolerance, corresponding to an RMS charge-density residual $\epsilon_\rho$ below $2\times10^{-3}$ within the first outer loop iteration. The DFT+QDMFT calculation using a three-bath discretization and ideal QAIMS simulations yields $\epsilon_\rho=5.60\times10^{-5}$, well below the prescribed tolerance and comparable to the CTQMC result of $\epsilon_\rho=3.19\times10^{-5}$. Increasing the bath size to five sites raises the residual to $\epsilon_\rho=3.77\times10^{-4}$, approximately one order of magnitude higher than for the three-bath case. This increase can be associated with the reduced ground-state fidelity achieved by QAIMS in the five-bath case, and the truncation of the continued fraction. Including sampling noise further increases the residuals to $\epsilon_\rho=6.97\times10^{-4}$ and $1.65\times10^{-3}$ for three and five bath sites, respectively. These results illustrate the impact of sampling noise on charge self-consistency. The RMS values are listed in Table~\ref{tab:rms_rho}.
\begin{table}[!b]
    \centering
    \begin{tabular}{@{}c|c@{}}
        \hline
        Solver & $\epsilon_{\rho}$ \\
        \hline
        CTQMC & $3.19\times 10^{-5}$ \\
        QAIMS ($N_b$=3, $N_k$=6, Ideal) & $5.60\times 10^{-5}$ \\
        QAIMS ($N_b$=3, $N_k$=6, Sampled) & $6.97\times 10^{-4}$ \\
        QAIMS ($N_b$=5, $N_k$=8, Ideal) & $3.77\times 10^{-4}$ \\
        QAIMS ($N_b$=5, $N_k$=8, Sampled) & $1.65\times 10^{-3}$ \\
        \hline
    \end{tabular}
    \caption{Converged RMS charge-density residuals for BCC lithium self-consistent calculations. Ideal QAIMS simulations solving three-bath AIMs yield a residual comparable to that obtained with CTQMC. Increasing the bath size to five sites raises the residual by approximately one order of magnitude. For both bath sizes, considering sampling increases the residual by approximately one order of magnitude.}
    \label{tab:rms_rho}
\end{table}

\begin{figure*}[!t]
    \centering
    \includegraphics[width=0.95\linewidth]{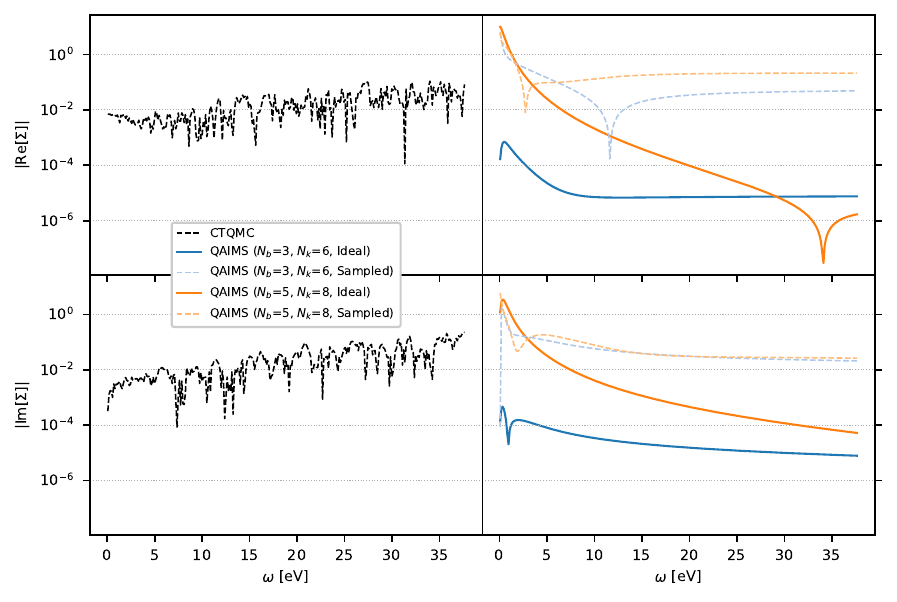}
    \caption{Absolute values of the real (top) and imaginary (bottom) parts of the impurity self-energy for BCC lithium. The left panels show the CTQMC reference, and the right panels show QAIMS results. Solid and dashed lines correspond to QAIMS simulations without and with sampling noise, respectively.}
    \label{fig:sigma_li}
\end{figure*}

The impurity self-energy provides a separate measure of the errors introduced by the approximate impurity solution. In the noninteracting limit, the exact self-energy vanishes, so any residual contribution tests the accuracy of the solver within the lattice calculation. Figure~\ref{fig:sigma_li} compares the absolute values of the real and imaginary parts of the impurity self-energy obtained with QAIMS and the reference CTQMC solver in their last inner loop iteration.

The DFT+QDMFT calculation using a three-bath discretization and ideal QAIMS simulations (solid blue lines) yields a smaller residual self-energy than the CTQMC reference (dashed black lines) over the plotted frequency range. For a five-bath discretization, QAIMS results from ideal simulations, shown as solid orange lines, are comparable to CTQMC over most of the frequency domain but exhibit larger deviations near zero frequency. This reduced accuracy is consistent with the lower ground-state fidelity obtained for the five-bath AIMs, as discussed in Section~\ref{sec:GS_fidelity}, as well as with the truncation of the Lanczos continued fraction at the eighth Krylov state, which may limit the accuracy of the impurity Green's function.

Sampling noise increases the magnitude of the impurity self-energy for both bath sizes. Errors in the measured and predicted moments, arising from reduced ground-state fidelities and the finite number of measurement shots, propagate through the impurity Green's function and its inversion via Dyson's equation, leading to deviations in the resulting self-energy. Despite these errors causing the inner QDMFT loop not to converge, the DFT+QDMFT calculations reach the prescribed charge-density convergence tolerance for both bath discretizations. These results demonstrate the feasibility of DFT+QDMFT calculations with relatively small bath discretizations, while highlighting the need to further improve the accuracy and reliability of QAIMS, or other quantum solvers, to address the increased computational and measurement challenges associated with larger bath models.

\section{Conclusions}
\label{sec:conclusions}

In this work, we improve our previous quantum impurity solver \cite{Karabin2026QDMFT} and integrate it into a self-consistent framework for DFT+QDMFT calculations. The newly implemented adaptive routines consistently identify the particle-number sector containing the ground state and, in most cases, provide accurate approximations to the AIM ground state and its particle and hole excitations throughout the inner QDMFT loop. To reduce measurement costs, we trained a ML model that predicts higher-order moments from low-order ones with an absolute relative error below one percent in 60 percent of the cases. These capabilities enable the application of QAIMS within a self-consistent electronic-structure workflow.

To demonstrate the framework we performed BCC lithium DFT+QDMFT calculations using three and five bath sites to discretize the hybridization function. For the larger bath, the discrete hybridization achieved RMS errors of order $\mathcal{O}(10^{-2})$, indicating a reasonable representation of the continuous bath. Most importantly, the DFT+QDMFT calculations reach charge self-consistency even in the presence of sampling noise, with RMS charge-density residuals below the prescribed tolerance of $2\times10^{-3}$. These results establish the feasibility of combining QAIMS with charge-self-consistent calculations when using relatively compact bath discretizations.

Extending this approach to larger bath sizes will require further improvements in the accuracy of the ground-state approximations, reductions in the measurement costs of the adaptive ground-state preparation and Hamiltonian moment estimation, and more capable ML models for predicting higher-order moments. These developments would allow the Lanczos continued fraction to be extended beyond the eight Krylov states considered here, paving the way for applications to larger and more strongly correlated materials.

\section{Acknowledgments}

The authors thank S. Kim and I.-S. Sue for useful discussions. 
This research has used resources of the Oak Ridge Leadership Computing Facility at the Oak Ridge National Laboratory, which is supported by the Office of Science of the U.S. Department of Energy under Contract No. DE-AC05-00OR22725. MK and HT is supported by the National Science Foundation under Grant No. 2538963. HT is supported by the National Science Foundation under Grant No. QIS-2328752. 

\bibliographystyle{unsrtnat}
\bibliography{references}

\end{document}